\documentstyle[twocolumn,aps,floats,babel,spanish,prl,latexsym,amsmath,amssymb,eqsecnum]{revtex}

\begin{document}

\draft \tolerance = 1000

\setcounter{topnumber}{1}
\renewcommand{\topfraction}{1}
\renewcommand{\textfraction}{0}
\renewcommand{\floatpagefraction}{1}
\newcommand{\br}{{\bf r}}

\twocolumn[\hsize\textwidth\columnwidth\hsize\csname@twocolumnfalse\endcsname

\title{{\rule{17.5cm}{.15mm}{\bfseries{\\\vspace{3mm}THE SPEED OF LIGHT AND THE FINE STRUCTURE CONSTANT\\\rule{17.5cm}{.15mm}}}}}
\author{Antonio Alfonso-Faus}
\address{E.U.I.T. Aeron\'autica\\
Plaza Cardenal Cisneros s/n  ~ Madrid 28040 ~ SPAIN\\
E-mail:aalfonso@euita.upm.es}
\maketitle
\vspace*{0.2cm}
\centerline{\em Accepted for publication in ``Physics Essays". Vol. \textbf{13} Nº1 March 2001}
\vspace*{0.cm}
\begin{abstract}
The fine structure constant $\alpha $ includes the speed of light as given
by $\alpha =\frac{e^{2}}{4\pi \varepsilon _{0}c\hbar }$. It is shown here
that, following a $TH\varepsilon \mu $ formalism, interpreting the
permittivity $\epsilon _{0}$ and permeabiliy $\mu _{0}$ of free space under
Lorentz local and position invariance, this is not the case. The result is a
new expression as $\alpha =\frac{e^{2}}{4\pi \hbar }$ in a new system of
units for the charge that preserves local and position invariance. Hence,
the speed of light does not explicitly enter in the constitution of the fine
structure constant. The new expressions for the Maxwell's equations are
derived and some cosmological implications discussed.\bigskip 

La constant de la structure fine ins\'{e}re aussi la
vitesse de la lumi\`{e}re en accord avec la formula $\alpha =\frac{e^{2}}{%
4\pi \varepsilon _{0}c\hbar }$. On d\'{e}montre avec ce travail que, suivant
le formulisme $TH\varepsilon \mu $ et interpr\'{e}tant la permitivit\'{e} $%
\varepsilon _{0}$ et la p\'{e}rmeabilit\'{e} $\mu _{0}$ du vide selon
l'invariant de Lorentz local et de position, cette formula n'est pas l'ad%
\'{e}quate. La nouvelle expression es $\alpha =\frac{e^{2}}{4\pi \hbar }$. 
dans un syst\`{e}me d'unit\'{e}s neuf pour la charge \'{e}lectrique, syst%
\`{e}me qui pr\'{e}serve l'invariant local et de position. Par cons\'{e}%
quent, la vitesse de la lumi\`{e}re ne rentre pas dans la constitution du
constant de la structure fine. On deduit les nouvelles expressions des \'{e}%
quations de Maxwell et on d\'{e}bat certaines inplications cosmologiques.\bigskip

\end{abstract}
\vspace{.5cm}

]

\section{\textbf{INTRODUCTION}}

The correct expression for the permittivity of free space $\varepsilon _{0}$%
, and consequently for the permeability of free space $\mu _{0}$, is of
fundamental importance when dealing with laboratory and cosmological work.
By correct we mean the forms that preserve both Local Lorentz Invariance
(LLI) and Local Position Invariance (LPI). We define both following Will
(see \cite{W}).\bigskip 

By LLI we mean the aspect of the Einstein's Equivalence Principle that
postulates the same predictions for identical local non-gravitational test
experiments, performed in two freely falling frames located at the same
event in space-time, but moving relative to each other.\bigskip

By LPI we mean the independence from the space-time location of the frame,
when observing the results of local non-gravitational test experiments.\bigskip

The $TH\varepsilon \mu $ formalism, devised by Lightman and Lee (\cite{LL}),
can be used to implement the Einstein's Equivalence Principle, as presented
by Will (\cite{W}). This author proves that a necessary and sufficient
condition for both Local Lorentz and Position Invariance to be valid is
given by

\begin{equation}
\varepsilon _{0}=\mu _{0}=\left( H_{0}/T_{0}\right) ^{1/2}  \label{rel1}
\end{equation}
for all events.

In our work here we use this relation to obtain Local Lorentz and Position
Invariance. Since the product $\varepsilon _{0}\mu _{0}$ is equal to $c^{-2} 
$, then the relation (\ref{rel1}) implies 
\begin{equation}
\varepsilon _{0}=\mu _{0}=\frac{1}{c}  \label{rel2}
\end{equation}

It is extremely important that we retain $c$ different from unity in all
equations. By doing so we keep open to a possible variation of $c$ with
cosmological time as a result of the expansion of the Universe, e.g.
Alfonso-Faus (\cite{A}), or any other cause. We are not dealing here with
the constancy or time variation of $c$. But to be able to find the correct
basic physical relations, in particular as related to the fine structure
constant, we must keep $c$ in the expressions.\bigskip

We know that in the electrostatic system of units (e.g. Jackson (\cite{J}))
one takes $\varepsilon _{0}=1$, and $\mu _{0}$ $=c^{-2}$ . In the
electromagnetic system of units $\epsilon _{0}$ $=c^{-2}$ and $\mu _{0}=1$.
Both are wrong from an invariance point of view, and the same happens with
the rationalized mks system. In the gaussian and Heaviside-Lorentz systems
one takes $\varepsilon _{0}=$ $\mu _{0}$ $=1$ which corresponds to take $c=1$. Here we are presenting a system of electromagnetic units consistent with
local and position invariance even in the case of a time varying $c$. By
doing so we keep the physical insight of the presence of $c$ in
electromagnetic properties.\bigskip

One of the results is that the speed of light does not enter explicitly in
the intrinsic constitution of the fine structure constant.\bigskip

On the other hand, we know that the constancy of the ratio of magnetic to
electric forces, as evidenced in the spectra from different distant
galaxies, implies that the fine structure constant must be a universal
constant. Once we prove that the speed of light does not enter in its
constitution, we get as a universal constant the ratio $e^{2}/\hbar $.

\section{\textbf{THE INVARIANCE OF THE MAXWELL EQUATIONS.}}

We follow Jackson (\cite{J}) for discussing units and dimensions of
electromagnetism. We choose length, mass, and time as independent, basic
units. The dimension of the ratio of charge and current will be that of
time. Then, the continuity equation for charge and current densities is 
\begin{equation}
\nabla \cdot \overrightarrow{J}+\frac{\partial \rho }{\partial t}=0
\label{rel3}
\end{equation}

We consider electromagnetic phenomena in free space, apart from the presence
of charges and currents. We have Coulomb's law

\begin{equation}
F_{1}=k_{1}\frac{qq^{\prime }}{r^{2}}  \label{rel4}
\end{equation}
and define the electric field as

\begin{equation}
E=k_{1}\frac{q}{r^{2}}  \label{rel5}
\end{equation}
where $k_{1}$ is a proportionality constant except for possible cosmological
time variations. We will call these ``constants'' by the name of factors.
Ampere's law giving the force per unit length between two infinitely long
parallel wires, separated by a distance $d$ and with currents $I$ and $%
I^{\prime }$ is 
\begin{equation}
\frac{dF_{2}}{dl}=2k_{2}\frac{II^{\prime }}{d}  \label{rel6}
\end{equation}
where $k_{2}$ is a factor as in (\ref{rel5}).\bigskip

By comparison of the magnitudes of the two mechanical forces (\ref{rel4})
and (\ref{rel6}) one has in free space

\begin{equation}
\frac{k_{1}}{k_{2}}=c^{2}  \label{rel7}
\end{equation}

We define the magnetic induction $B$ from Ampere's law for a long straight
wire carrying a current $I$,

\begin{equation}
B=2k_{2}\frac{\beta I}{d}  \label{rel8}
\end{equation}
where $\beta $ is a proportionality factor, which has dimensions. Combining (%
\ref{rel3}), (\ref{rel5}), (\ref{rel7}) and (\ref{rel8}) one has the
dimensions of $E/B$ to be $l/(t\beta )$. As we will see $E$ and $B$ will
have the same dimensions, so that $\beta $ has the dimensions of velocity.\bigskip

Finally, Faraday's law of induction gives

\begin{equation}
\nabla \times \overrightarrow{E}+k_{3}\frac{\partial \overrightarrow{B}}{%
\partial t}=0  \label{rel9}
\end{equation}
On the basis of Galilean invariance one has $k_{3}=1/\beta $. But it can
also be proved from the Maxwell 's equations using the fields already
defined and taking into account that in the wave equation the velocity of
propagation is the speed of light. Maxwell's equations are then 
\begin{eqnarray}
\nabla \cdot \overrightarrow{E} &=&4\pi k_{1}\rho  \label{rel10} \\
\nabla \times \overrightarrow{E}&=&-k_{3}\frac{\partial \overrightarrow{B}}{%
\partial t}  \notag \\
\nabla \times \overrightarrow{B} &=&4\pi k_{1}\beta \overrightarrow{J}%
+(k_{2}/k_{1})\beta \frac{\partial \overrightarrow{E}}{\partial t}  \notag
\\
\nabla \cdot \overrightarrow{B} &=&0  \notag
\end{eqnarray}
where we have the relations between the proportionality factors:

\begin{eqnarray}
k_{1}/k_{2} &=&c^{2}  \label{rel11} \\
k_{3}\beta &=&1  \notag
\end{eqnarray}
We only have two factors that must be chosen to completely define the
system. We will do it keeping the LLI and the LPT. Also, from the Lorentz
force expressed in the gaussian or Heaviside-Lorentz Systems that have $%
\varepsilon _{0}$ and $\mu _{0}$ equal, 
\begin{equation}
\overrightarrow{F}=q(\overrightarrow{E}+\frac{\overrightarrow{v}}{c}\times 
\overrightarrow{B})  \label{rel12}
\end{equation}
hence $E$ and $B$ have the same dimensions. For electromagnetic waves in
free space $E=B$ and we keep this result, so that from (\ref{rel9}) one has 
\begin{equation}
k_{3}=c^{-1}  \label{rel13}
\end{equation}

So far we have kept the same classical discussion as in Jackson (\cite{J}). We are
left with only one factor to be defined and this is the characteristic step
of this work. From the LLI and the LPI one has the factor in Coulomb's law: 
\begin{equation}
k_{1}=c  \label{rel14}
\end{equation}
and therefore 
\begin{equation}
k_{2}=c^{-1}  \label{rel15}
\end{equation}

The laws of electromagnetism are then given in LLI and LPT form as follows:

Coulomb's law: 
\begin{equation}
F_{1}=c\frac{qq^{\prime }}{r^{2}}  \label{rel16}
\end{equation}

Electric field 
\begin{equation}
E=c\frac{q}{r^{2}}  \label{rel17}
\end{equation}

Ampere's law 
\begin{equation}
\frac{dF_{2}}{dl}=\frac{2}{c}\frac{II^{\prime }}{d}  \label{rel18}
\end{equation}

Magnetic induction 
\begin{equation}
B=\frac{2I}{d}  \label{rel19}
\end{equation}
and finally Maxwell 's equations are in this form: 
\begin{eqnarray}
\nabla \cdot \overrightarrow{E} &=&4\pi c\rho  \label{rel20} \\
\nabla \times \overrightarrow{E}&=-&\frac{1}{c}\frac{\partial \overrightarrow{B%
}}{\partial t}   \notag \\
\nabla \times \overrightarrow{B} &=&4\pi \overrightarrow{J}+\frac{1}{c}%
\frac{\partial \overrightarrow{E}}{\partial t}  \notag \\
\nabla \cdot \overrightarrow{B} &=&0  \notag
\end{eqnarray}

\section{\textbf{COSMOLOGICAL IMPLICATIONS}}

The fine structure constant $\alpha $ must be a real constant as inferred
from the observations of spectra from different distant sources through the
ratio of magnetic spin-orbit interaction to the electric interaction. The
possible variation reported by Webb et al. (\cite{WE}) is very small for a
cosmological time variation at a distance of about red shift one. They
report a change of about one part in $10^{5}$, while here we are considering
variations of the same order as the value of the constant. The same
consideration applies to the reported result of the electroweak measurement
of the strengthening of the electromagnetic coupling (Levine et al. (\cite{L}%
)): a value of $128.5$ as compared with $137.0$ represents a small variation
in cosmological terms.

Let us assume that $\alpha $ is a constant. We have proved that it only
depends on $\hbar $ and $e$. Let us assume that $\hbar $ and therefore $e$
are also constants. In terms of the Rydberg constant $R_{\infty }$ and the
Bohr radius $a_{0}$ one has the following relation (see \cite{E}) : 
\begin{eqnarray}
R_{\infty }hc &=&13.6056981eV=m_{e}c^{2}\alpha ^{2}  \label{rel21} \\
2R_{\infty }hc &=&\frac{e^{2}}{4\pi \varepsilon _{0}a_{0}}=27.2113961eV 
\notag
\end{eqnarray}

Here $h$ is $2\pi \hbar $. If we take this energy as constant then a time
variation in $c$ implies a time variation in the mass of the electron. Also,
taking $h$ and $e$ constant one has from (\ref{rel21}) 
\begin{eqnarray}
R_{\infty }c &=&const.  \label{rel22} \\
\varepsilon _{0}a_{0} &=&const.  \notag
\end{eqnarray}
Since $\varepsilon _{0}=c^{-1}$ one has the result 
\begin{equation}
a_{0}\propto c  \label{rel23}
\end{equation}

The conclusion is that a time variation in the speed of light, keeping $%
\alpha $, $\hbar $ and $e$ constants imply that the masses and sizes of the
quantum world vary also with time. In particular, if the speed of light
decreases with time due to the expansion of the Universe (Alfonso-Faus (\cite
{A})), the sizes of the quantum particles also decrease. Both effects would
go together: the expansion of the Universe and the contraction of the
quantum world at the same time.\bigskip

\section{\textbf{CONCLUSION}}

We have reviewed all basic electromagnetic relations with the choice $%
\varepsilon _{0}=\mu _{0}=1/c$ that preserves local Lorentz invariance and
local position invariance. One immediate consequence is that there is a
factor of $c$ in Coulomb's law multiplying the product of the two charges.
In this new system of units the speed of light is explicit in the places
required by the invariance condition imposed, which gives an insight in the
intrinsic relations of physical properties. In particular, the fine
structure constant given as $\frac{e^{2}}{4\pi \varepsilon _{0}c\hbar }$ is
now, in the new system, $\frac{e^{2}}{4\pi \hbar }$ . Hence, the constancy
of the fine structure constant, deduced by means of observations of spectra
from different distant sources through the ratio of magnetic spin-orbit
interaction to the electric interaction, is in fact the constancy of $\frac{%
e^{2}}{\hbar }$ . This is very important for theories that contemplate
time-varying physical ''constants'' (e.g. Alfonso-Faus (\cite{A})). In
particular, the exact constancy of the electric charge of the electron would
imply the exact constancy of Planck's constant. If there is no cosmological
effect on $e$, then there is no cosmological effect on $\hbar $, regardless
of any possible cosmological effect on $c$. If there is a cosmological
effect on $c$ then there is a cosmological effect on the masses and sizes of
quantum particles, keeping constant $\hbar ,e$ and the atomic energy
levels.\bigskip 

\textbf{{\large {\Large A}}CKNOWLEDGEMENTS}

I am indebted to Prof. A. Estevez and Prof. A. Alcazar de Velasco for
reviewing the manuscript. I am also grateful to Dr. T. LeCompte and Mr. F.
E. Reed for usefull discussions.

\end{document}